\documentclass[rnote]{aa} % for the research notes
\usepackage{graphicx}
\usepackage{color}

\usepackage{txfonts}
\begin{document}

   \title{Revealing the nature of star forming blue early-type galaxies at low redshift \thanks{Based on data compiled from Galaxy Zoo project, and the volunteers 
contribution are acknowledged at http://www.galaxyzoo.org/Volunteers.aspx} }

 %  \subtitle{I. Overviewing the $\kappa$-mechanism}

   \author{Koshy George
          \inst{1}
          \and
          Kshama Zingade\inst{2}}

   \institute{Indian Institute of Astrophysics, 2nd Block, Koramangala, 
Bangalore - 560034, India\\
              \email{koshy@iiap.res.in}
         \and
             Indian Institute of Technology Madras, IIT P.O, Chennai 600 036
India\\
             \email{kshamazings@gmail.com}}

%\thanks{Based on data compiled from Galaxy Zoo project, and the volunteers contribution are acknowledged at http://www.galaxyzoo.org/Volunteers.aspx}\\

%   \date{Received September 15, 1996; accepted March 16, 1997}

% \abstract{}{}{}{}{} 
% 5 {} token are mandatory
 
  \abstract
   {Star forming early-type galaxies with blue optical colours at low redshift can be used to test our current understanding of galaxy formation and evolution.}
   {We want to reveal the fuel and triggering mechanism for star formation in these 
otherwise passively evolving red and dead stellar systems.}
   {We undertook an optical and ultraviolet study of 55 star forming blue early-type galaxies,
 searching for signatures of recent interactions that could be driving the molecular gas into the galaxy 
and potentially triggering the star formation.}
   {We report here our results on star forming blue early-type galaxies with tidal trails and in close proximity to neighbouring galaxies that are evidence of ongoing or recent interactions between galaxies. There are 12 galaxies with close companions with similar redshifts, among which 
two galaxies are having ongoing interactions that potentially trigger the star formation. Two galaxies show a jet feature that could be due to the 
complete tidal disruption of the companion galaxy. The interacting galaxies have high star formation rates and very blue optical colours. Galaxies with no 
companion could have undergone a minor merger in the recent past.}
   {The recent or ongoing interaction with a gas-rich neigbouring galaxy could be responsible for bringing cold gas 
to an otherwise passively evolving early-type galaxy. The sudden gas supply could trigger the star formation, eventually creating a 
blue early-type galaxy. The galaxies with ongoing tidal interaction are blue and star forming, thereby implying that blue early-type galaxies can exist even when the companion is on flyby so does not end up in a merger.}

%The fuel and trigger for star formation could be driven by gravitational interaction or mergers between neighbouring galaxies. Blue ETGs 
%could be a transient phase for an otherwise passively evolving red ETGs. }

   \keywords{galaxy evolution, elliptical and lenticular, galaxy interactions, galaxy star formation
               }

\maketitle
%
%________________________________________________________________

\section{Introduction}

Early-type galaxies (ETGs) are morphologically classified as ellipticals or lenticulars based on visual inspection of imaging data. They are observed to be passively evolving stellar systems that 
follow a  tight red sequence when placed on the galaxy colour magnitude diagram \citep{Faber_1973,Visvanathan_77,Baldry_2004}. These galaxies are predominant in dense environments like 
the centres of galaxy clusters, which are hostile to 
star formation \citep{Dressler_1987}. This observational evidence makes them classic example of systems that formed at very early epochs ($z>2$), followed by passive evolution  \citep{Renzini06}. However, there is
growing evidence from deep surveys in recent years that have demonstrated that the stellar mass in red sequence galaxies has been increasing over the past eight billion years \citep{Bell_2004,Faber_2007,Brown_2007}. 
 This could happen because of stellar mass growth through scenarios such as galaxy mergers, tidal interactions, and in situ star formation. These processes operate in the field, increasing
 the stellar mass of ETGs by accreting more stars and gas, thereby triggering star formation throughout an existing base population of evolved stars. Star formation can 
then elevate the blue spectral energy distribution of ETGs, which will change 
their position from the red sequence to the blue cloud on the galaxy colour-magnitude diagram. 

Indeed, a sample of such ETGs have been observed to be significantly bluer than the red sequence galaxies (\citealt{schawinski09}, hereafter S09; \citealt{McIntosh14}).
 S09 have reported 204 blue ETGs at low redshifts (0.02 $<$ $z$ $<$ 0.05), with approximately L$*$ luminosities. The early-type morphology of these galaxies 
has been confirmed based on visual classifications from the galaxy zoo. They are observed to have a range of velocity dispersions ($\sigma <$ 200 km/s), as 
found in low density environments and make up 5.7 $\pm$ 0.4 $\%$ of the low-redshift early-type galaxy population. Blue ETGs are absent above  $u-r$ colour $\sim$ 2.5 mag, a region that is populated by
 red sequence galaxies on the colour-magnitude diagram [S09]. S09 have classified the blue ETGs based on the positions on line diagnostic diagram \citep{Baldwin_1981} with 25 $\%$ as only star forming, 25 $\%$ as both star 
forming and AGN, 12 $\%$ as AGN, and 38 $\%$ as having no strong emission lines to classify. The star formation rates of the 55 star forming blue ETGs are very high (0.45 to 21 M$_{\odot}$/yr) 
and the colours are too blue for any such systems found in the local universe. We are interested in revealing the mechanisms responsible for star formation in the blue ETGs 
as it helps in our understanding of galaxy formation and evolution. We speculate a  cold gas supply either from recent gas- rich merger or tidal interaction with a neighbouring 
gas-rich galaxy, responsible for the observed intense burst of star formation. We report here our initial results based on an optical and ultraviolet imaging analysis of 55 star forming blue ETGs.
We adopt a flat Universe cosmology with 
%$H_{\rmn{o}} = 71\,\mathrm{km\,s^{-1}\,Mpc^{-1}}$, $\Omega_{\rmn{M}} = 0.27$, $\Omega_{\Lambda} = 0.73$ \citep{Komatsu_2011}.\\
$H_{o} = 71\,\mathrm{km\,s^{-1}\,Mpc^{-1}}$, $\Omega_{M} = 0.27$, $\Omega_{\Lambda} = 0.73$ \citep{Komatsu_2011}.\\

\section{Data and analysis}

\begin{figure}
\resizebox{\hsize}{!}
{\includegraphics{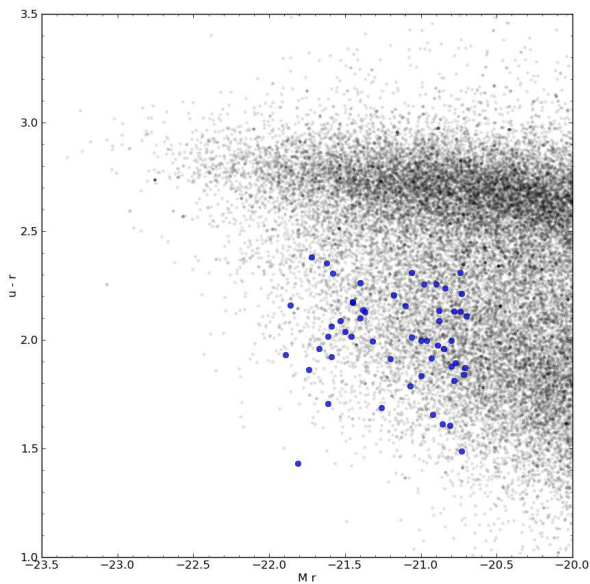}}
\caption{Colour-magnitude diagram of galaxies in the redshift range of 0.02 < z < 0.05. The 55 star forming 
blue ETGs (blue) are overplotted. Early-type galaxies form a tight red sequence, and the late-type galaxies are scattered in the  blue cloud on the colour magnitude diagram. }
\label{fig:fig1}
\end{figure}

We used SDSS DR7 \citep{Abazajian_2009} $u$ and $r$ band photometric data,  $g$  band imaging data, and $gri$ composite images for our analysis. The near ultraviolet galaxy images were taken from the GALEX 
(Galaxy Evolution Explorer) database \citep{Martin05}. We constructed the colour magnitude diagram for  galaxies in the redshift range of 0.02 < z < 0.05 using the data generated from a volume-limited 
catalogue of SDSS galaxies with photometric errors less than 0.2 mag.
The galaxy's absolute magnitudes are based on SDSS $petromag,$ and the colours are derived from SDSS $modelMags$. Figure~\ref{fig:fig1} shows the colour-magnitude diagram with a blue cloud 
populated by late-type galaxies and a red-sequence dominated by ETGs. The 55 ETGs studied in this work fall in the blue cloud irrespective of their
 early-type morphological classification.\\

%$\sim$ 74000%

\subsection{Optical imaging analysis}
\begin{figure}
%\begin{center}
\resizebox{\hsize}{!}
{\includegraphics{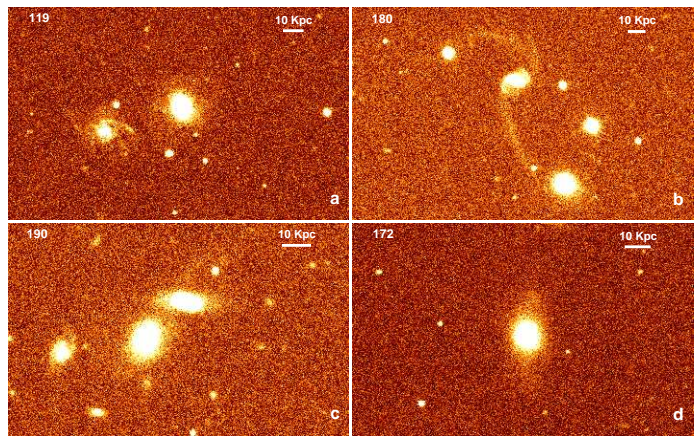}}
\caption{Blue ETGs SDSS $g$ band (colour-enhanced) images. Representative sample of galaxies with disturbed neighbours (a), interacting systems (b,c), and galaxies with a jet feature (d) are shown. The object ID from S09
 is assigned for the blue ETGs in the images. Images are zoomed and scaled to different levels to highlight the interaction features. These images are taken with a 54$s$ integration time typical of 
SDSS imaging data.}
\label{fig:fig2}
 \end{figure}

We analysed 55 SDSS $gri$ composite images of  $\sim$ 6'$\times$6' field centred on the galaxy to check for any possible close encounters with neighbouring galaxies. The SDSS images have a plate scale 
of 0.396~$\arcsec$/pixel, and the chosen field 
corresponds to a physical size of $\sim $ 144 Kpc $\times$ 144 Kpc at $z$ $\sim$ 0.02. The region we selected encompasses the virial radius of a typical L$*$ galaxy, a criterion that is  crucial for studies of the
 environments of blue ETGs. The companion galaxies in the
 $\sim$ 6'$\times$6' field are checked for their proximities, spectroscopic redshifts, and possible 
interactions with the blue early-type galaxy. Companion galaxies within the chosen field that share a 
minimum redshift increment of $\delta z < 0.002$  ($\delta$ v < 600 Km/s) were classified as neighbours. Based on this criterion, we found a 
total of twelve blue ETGs with neighbouring galaxies, and this corresponds to $\sim$ 22 $\%$ of the star forming blue ETG sample. The $g$ band images of these galaxies are then analysed for 
the signs of interactions and the projected distance between the galaxy and the immediate neighbour. The blue ETGs studied here show 
signatures of recent or ongoing interactions. Two galaxies are found to have ongoing tidal interactions with trails through which 
stellar material is connected. We could also identify two galaxies with an 
extended jet feature. Four galaxies are found to have companions with a disturbed morphology that can occur because of the tidal interaction with the blue ETG. The details of twelve blue ETGs with
 neighbouring companions and of two galaxies having unusual features are given in Table~\ref{galaxy details}. We adopted the galaxy-naming convention following the S09 objID.
 Figure~\ref{fig:fig2} shows the images of the representative sample from our analysis.\\

\begin{table*}
\caption{\label{t7}Blue ETGs with ongoing tidal interactions and/or observed to have a close companion with equivalent redshifts.}
\label{galaxy details} 
\tiny
\centering
\begin{tabular}{cccccccccl}
\hline\hline
%SDSS & S09  &  RA & DEC  & $z$  &   M$_r$ &  $u-r$ &  SSFR &  $z_$neighbour & Distance to   &  Details of \\ 
SDSS & S09 &  RA & DEC  & $z$  &   M$_r$ &  $u-r$ &  SSFR &  $z_{comp}$    &  companion \\ 
%objID  &  h:m:s  & d:m:s &  &  (mag)    &  (mag)    &  (M$\odot$/yr) &  neighbour & companion (kpc) & companion\\
ID & objID &  h:m:s  & d:m:s &  &  (mag)    &  (mag)    &  (yr$^{-1}$) &    & galaxy \\
\hline
587725550139408424 & 14  &12:35:02.6&    $+$66:22:33.4&   0.047& $-$21.67 &  1.959 &      -10.25  & 0.047&blue dwarf (strong H$\alpha$ emission line)\\
587726100953432183 & 30  &15:17:19.7&    $+$03:19:18.9&   0.037& $-$20.74 &  2.131 &      -10.75  & 0.037&red ETG\\
587730816286785593 & 61  &22:15:16.2&    $-$09:15:47.6&   0.038& $-$21.61 &  1.707 &       -9.64  & --  &jet feature\\
587731500261834946 & 66  &10:54:37.9&    $+$55:39:46.0&   0.048& $-$20.89 &  1.976 &       -9.95  & 0.048&red ETG\\
587732769982906490 & 92  &12:20:23.1&    $+$08:51:37.1&   0.049& $-$20.88 &  2.086 &      -10.23  & 0.049&blue disturbed disc (strong H$\alpha$ emission line)\\
587733412064788620 & 103 &15:50:00.5&    $+$41:58:11.2&   0.034& $-$20.80 &  1.879 &      -10.18  & 0.035&red ETG\\
587735742615388296 & 119 &15:53:35.6&    $+$32:18:20.6&   0.050& $-$21.07 &  1.789 &      -10.07  & 0.050&blue disturbed spiral galaxy (strong H$\alpha$ emission line)\\
587739115234394179 & 139 &07:56:08.7&    $+$17:22:50.5&   0.030& $-$20.73 &  2.211 &      -10.62  & 0.028&blue disturbed galaxy (strong H$\alpha$ emission line)\\
587741490904105107 & 160 &10:25:24.7&    $+$27:25:06.3&   0.050& $-$20.98 &  2.256 &      -10.42  & 0.049&red ETG\\
588007005234856197 & 172 &08:17:56.3&    $+$47:07:19.5&   0.039& $-$21.06 &  2.309 &      -10.11  & --  &jet feature\\
588011125186691149 & 180 &12:06:17.0&    $+$63:38:19.0&   0.040& $-$21.26 &  1.686 &       -9.59  & 0.040& blue spiral galaxy interacting (strong H$\alpha$ emission line)\\
588017604156784724 & 190 &14:14:33.2&    $+$40:45:22.9&   0.042& $-$20.86 &  1.614 &       -9.92  & 0.041&blue galaxy interacting (strong H$\alpha$ emission line)\\
588017991233110227 & 206 &14:37:33.0&    $+$08:04:43.0&   0.050& $-$21.62 &  2.353 &      -10.55  & 0.051&dwarf galaxy\\
588018254297038899 & 209 &16:18:18.7&    $+$34:06:40.1&   0.047& $-$21.58 &  2.307 &      -10.59  & 0.048&blue disturbed spiral galaxy (strong H$\alpha$ emission line)\\
\hline
\end{tabular}
\tablefoot{Column (1) is the SDSS DR7 ObjID; columns (2)  the object ID for the galaxies taken from S09; columns (3) and (4) galaxy coordinates (epoch J2000); column (5) the spectroscopic redshift of blue ETGs from SDSS; column (6) the  
$r$-band absolute magnitude; column (7) the $u-r$ colours; column (8) the specific star formation rates; column (9) the spectroscopic redshift ($z$comp) of nearby companions from SDSS; column(10) 
details obtained from the visual inspection of SDSS $gri$ composite images and the companion-galaxy SDSS spectrum.\\
}
\end{table*}

\subsection{Ultraviolet imaging analysis}

\begin{figure}
%\centering
\resizebox{\hsize}{!}
{\includegraphics{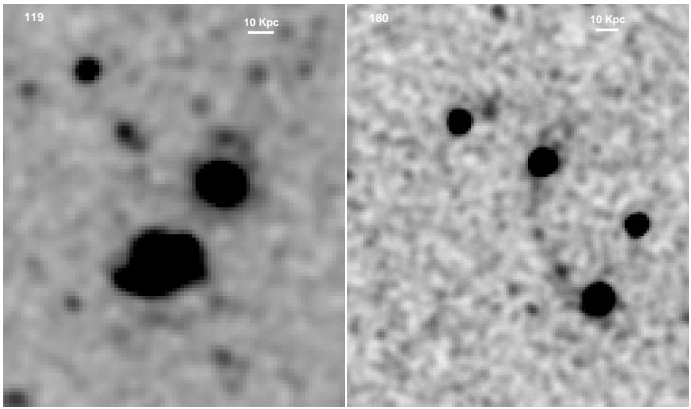}}
\caption{GALEX near-UV colour-inverted images of selected blue ETGs. (a) Galaxy 119 with integration time $\sim$ 3136$s;$ (b) Galaxy 180 with integration time 390$s$.}
\label{fig:fig3}
 \end{figure}

The ultraviolet (UV) spectral energy distribution of a star forming galaxy is dominated by the flux from the young, hot, massive stars. The 
near UV images of blue ETGs could effectively trace the signatures of ongoing star formation. 
We checked GALEX near-UV (1750-2750 $\AA$) database images of 12 galaxies with neighbours for possible interaction-induced star formation. The GALEX near-UV channel is dominated by the zodiacal
light with typical surface brightness levels of 26.5 mag/arcsec$^2$, and the GALEX plate scale (1.5 $\arcsec$/pixel) is coarser than that of the SDSS images \citep{Martin05}. 
This makes the detection of finer details in near-UV very difficult.    
There are GALEX data for 10 out of the 12 galaxies with neighbours, and one galaxy (190) with ongoing interaction has no GALEX observation.  
The contribution from the zodiacal light to the noise level of the images was suppressed by employing statistical smoothing techniques. We performed a Gaussian -smoothing using a kernel of a width of three pixels to enhance the signatures of star formation along the interacting trails.  
Figure~\ref{fig:fig3} shows the selected near-UV images of two galaxies. Galaxy 119 and the 
neighbouring galaxy are found to have enhanced emission in near-UV when compared to optical data. This could be due to the sudden burst 
of intense star formation due to the gravitational interaction between galaxies. We could not detect an 
interacting trail between
the galaxies in the near UV data. However, there can be an exchange of molecular gas (fueling star formation) between the galaxies, 
which is only detected from radio or sub-mm observations. Galaxy
 180, on the other hand, has interaction trails with the companion, and the blobs of
 enhanced emission are clearly visible in the near-UV images. This could be due to triggered star formation episodes happening within the 
interacting trails of the galaxies, or we are witnessing the burst of ongoing star formation in the arms 
of the spiral galaxy that is being stripped by the blue ETG.\\

\subsection{Genuine case of tidal interaction}

The genuine cases of tidal interactions are seen in Figure 2. Figure 2a shows Galaxy 119  with a blue companion 
that appears to be disrupted in morphology owing to the blue ETG. The companion galaxy is having ongoing star formation as is evident from the near-UV image (Figure 3a). 
Figure 2b shows Galaxy 180, which hosts a spiral companion with
disrupted morphology. The system is interacting, and the blue ETG have
 a very high star formation rate of 18 M$_{\odot}$/yr. The spiral companion is being stripped of its spiral arms with
tidal trails connecting the galaxies, which is the site of ongoing
star formation, as seen in the near-UV image (Figure 3b). Figure 2c shows Galaxy 190, which is interacting with a blue companion. 
Figure 2d shows Galaxy 172 with a jet feature that appears to be a tidally destroyed companion galaxy. We note that 
the SDSS exposure time (54s) is not adequate for bringing out the finer details, which can only be achieved with deep imaging data.\\

\section{Galaxy colour-magnitude and star formation diagrams}

\begin{figure}
\resizebox{\hsize}{!}
%\centering
{\includegraphics{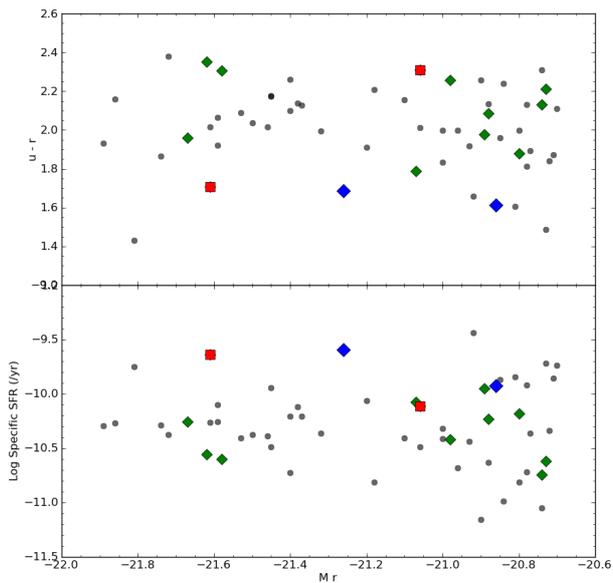}}
\caption{Galaxy r-band absolute magnitude vs $u-r$ colour and specific star formation rates. The star forming blue ETGs are shown in black, twelve galaxies with neighbours with green diamonds, interacting galaxies with blue diamonds, and 
the galaxies with jet features with a red square.}
\label{fig:fig4}
\end{figure}

We plotted 55 blue ETGs in the galaxy optical colour- vs absolute-magnitude diagram and in the absolute-magnitude vs specific star 
formation rate diagrams as shown in Figure~\ref{fig:fig4}. The aperture-corrected star formation rate (data from S09) is derived using the H$\alpha$ emission line strength measured from the  3$\arcsec$ SDSS fibre spectrum of 
the inner 3 Kpc (at $z \sim 0.05 $) of the galaxy (see Section 3.2 of S09 for details) and scaled for the galaxy-wide star formation. The specific star formation rates were computed using the 
galaxy stellar mass estimates from the best-fit spectral energy distribution using the flexible stellar population synthesis code \citep{Conroy_2009}. The colours were derived from SDSS $modelMags$, which
 measures the ratio of the global flux distribution of the galaxy observed in two different filter pass bands. Star formation affects both the $u-r$ colour and the 
absolute magnitude, making the galaxies appear both brighter and bluer on the colour-magnitude diagram. The 
galaxies on the magnitude star formation rate diagram show a trend toward brighter galaxies having 
high star formation rates. It is clear from these diagrams 
that the interacting galaxies have high star formation rates and very blue colours. This is 
a strong indication that interaction between galaxies triggers the star formation 
(We do caution that it is difficult
to make a general conclusion based on data from two 
interacting systems.) However, it is likely that other systems might have undergone
 such an interaction in the recent past. This is supported by the distribution of the 12 galaxies with close 
companions in the diagrams. Galaxies with a jet feature have high star formation rates though the 
colours are redder for one galaxy than in the rest of
the sample.\\

%{\includegraphics[width=10cm,height=10cm]{name}}

\section{Large scale environment analysis}

It has been known since the seminal work of \citet{Dressler_1980} that galaxies are prone to morphological transformation in a dense environment. This is supported by the observations of 
 the dominance of ellipticals and S0 galaxies in dense galaxy clusters compared to star forming spiral galaxies. Galaxies get transformed from spirals to ellipticals and S0s when they fall into dense environments. 
The blue early-type galaxies reported here could 
undergo preprocessing while in groups or outskirts of galaxy clusters. 

\begin{figure}
\resizebox{\hsize}{!}
%\centering
{\includegraphics{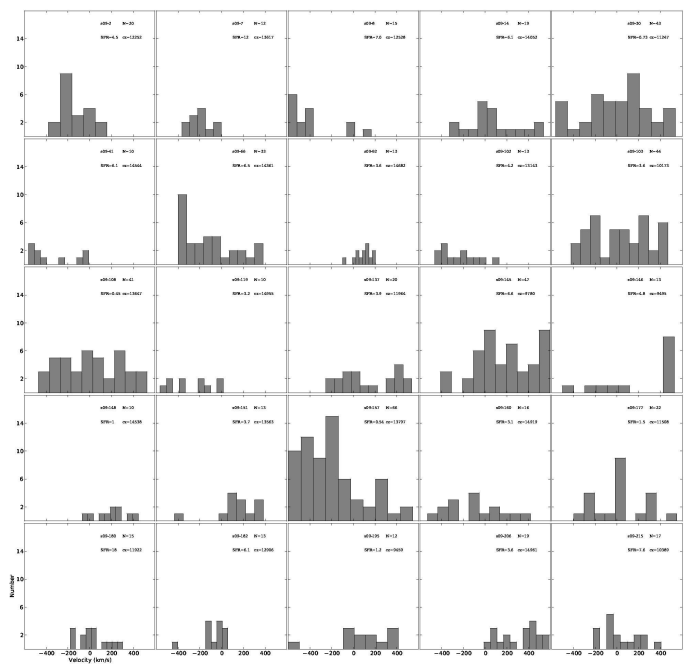}}
\caption{Histogram of galaxies within a 1 Mpc search radius of blue early-type galaxies. S09 ID, number of galaxies, star formation rate (M$_{\odot}$/yr), and radial velocity (Km/s) are noted 
on each histogram.}
\label{fig:fig5}
\end{figure}

The large scale environment of the blue early-type galaxies were analysed to check for an environmental dependence of
 the observed high star formation rates, whence the blue colours of these galaxies. Galaxies within a region 1 Mpc 
 centred on the cordinates of 55 star forming blue early-type galaxies were searched for the presence of galaxy groups or clusters. The NASA Extragalactic Database (NED) \footnote{https://ned.ipac.caltech.edu/} 
was used to get the data of neighbouring galaxies. Galaxies were selected above the SDSS limiting magnitude of  $g$ $\sim$ 21 and within radial velocity of 1200 km/s around the blue early-type galaxy. 
Figure~\ref{fig:fig5} highlights the results of the analysis as shown in the histogram of galaxies within a 1 Mpc search radius of the blue early-type galaxies studied here. Only galaxies with 
ten companions and more are shown here. Different bin sizes are chosen to make the histograms on a common velocity scale. Galaxies with high star formation rates 
 are found in varying environmental densities, implying that large scale environmental effects are not seen to enhance star formation in blue early-type galaxies.

\section{Radio data}

\begin{figure}
\resizebox{\hsize}{!}
%\centering
{\includegraphics{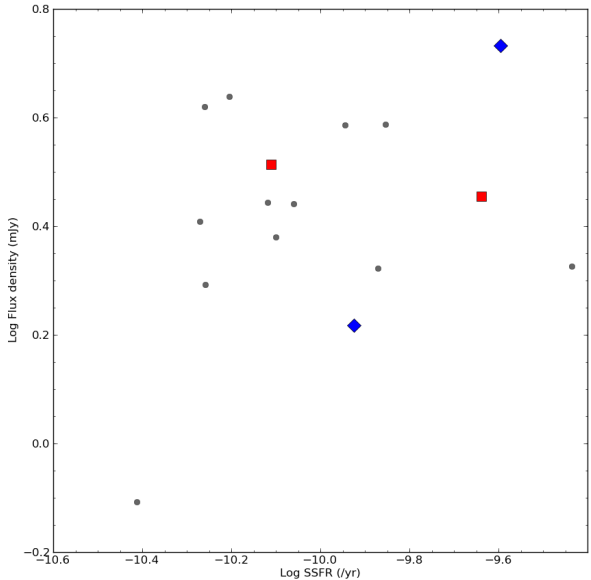}}
\caption{1.4 GHz continum flux density of blue early-type galaxies from FIRST survey catalogue plotted against the specific star formation rates. The star forming blue ETGs are shown in black, interacting galaxies in blue diamond, and 
the galaxies with jet features in the red square.}
\label{fig:fig6}
\end{figure}

The high star formation rates of blue early-type galaxies need an abundant supply of the cold molecular hydrogen that fuels them. In the absence of HI/CO data on these galaxies, we checked for other radio data that can be
 used as a proxy to trace the molecular hydrogen. Data on 16 blue early-type galaxies exist at 1.4 GHz from the FIRST (Faint Images of the Radio Sky at Twenty-cm) survey using Very Large Array \citep{Becker_95}. Radio sources 
within a separation of 10 arc sec from the galaxy optical coordinates are reported with flux density (in mJy) in the 1.4 GHz radio continum. There are 16 galaxies with 1.4 GHz continum 
flux density above the survey detection threshold of 1 mJy. Figure~\ref{fig:fig6} shows the plot of specific star formation rates against the detected 1.4 GHz continum flux. It is interesting to note
 that the galaxies with neighbours do not have flux detection, irrespective of the FIRST survey observations. The radio data can at least trace the star formation (if not the presence of neutral hydrogen) for the 16 
galaxies shown here. Future HI observations with VLA/GMRT should shed more light on the neutral hydrogen content and kinematics in blue early-type galaxies.

\section{Discussion}

The existence of morphologically disturbed and peculiar red sequence galaxies are studied in detail by \citet{Kaviraj_2011} and \citet{Kaviraj_2010}
 using SDSS and HST imaging data, respectively. These studies conclude 
that mergers are responsible for the morphological disturbance found in their sample of ETGs. The galaxies studied in our work are blue, and a few have ongoing interactions that 
make it an ideal case of `caught in the act'. That these galaxies are blue and star forming during the phase of interaction (galaxy 180 and 190), point towards a scenario in which
 such galaxies can exist without a merger scenario. The star formation rates of the blue ETGs are very high, making them brighter and bluer 
in comparison with the red sequence galaxies. The high star formation rates are only 
possible with an abundant supply of cold gas. Early-type galaxies are generally gas-poor systems, and they cannot support fuel for such high star formation rates and for the galaxy-scale blue colours 
found in blue ETGs \citep{Young_2011,Young_2014}. This prompted us to speculate that the fuel for star formation could be accreted from the companion galaxy, which could be a possible
 scenario at least in the case of galaxies where one can see ongoing interactions. The companion galaxies are 
significantly bluer in comparison with the blue ETGs. (Details of 
companion galaxies are given in Table~\ref{companion galaxy details}.) They could be potentially gas-rich galaxies and show signatures of active star formation 
(in terms of H$\alpha$ emission line strength) in the SDSS spectra. The existence of blue ETGs 
can be explanied by the frosting scenario of residual star formation found in local ETGs on the red sequence, in which an early-type galaxy may have 
triggered star formation over the old base population \citep{Trager2000,Serra_2007}. The initial phase of this scenario 
will rapidly move the galaxy from the red sequence to the blue cloud and will subsequently change the position on the galaxy colour magnitude diagram. These galaxies will 
have high star formation rates (depending on the availability of cold gas) and will move back to the red sequence within one billion years once the gas supply or 
star formation ceases. We speculate that the blue ETGs studied here are at different phases of getting cold gas from the companion and have migrated from the red sequence 
to the blue cloud. The interacting galaxies could be in the most efficient case of gas accretion, as is evident from the high star formation rate (18 M$_{\odot}$ /yr).
The wide range of star formation rates in these galaxies might correspond to the scenario where we are 
observing the galaxy at different periods of the star formation activity. The virialisation time scale for 
a L$*$ galaxy could be smaller than the time scale of ongoing star formation. This, along with the shallow SDSS imaging, explains the reason 
for not seeing interaction features in many of the star forming blue ETGs.\\

\begin{table}
\caption{Details of the companion galaxy $r$ band absolute magnitudes and colours and a rough estimate of the projected centre-to-centre distance (in Kpc) from the blue early-type galaxy.}              % title of Table
\label{companion galaxy details}      % is used to refer this table in the text
\centering                                      % used for centering table
\begin{tabular}{c c c c}          % centered columns (4 columns)
 \hline
\hline
S09  &  Comp M$_r$     &  Comp $u-r$ & Comp $\sim$ D \\ 
objID  &   (mag)    &  (mag)  & (Kpc)\\
\hline  
14&     -19.64  &1.23&  33\\
30&     -21.90  &2.93&  126\\
66&     -21.05  &2.63&  138\\
92&     -19.21  &1.10&  90\\
103&    -21.53  &2.77&  103\\
119&    -19.03  &1.00&  41\\
139&    -18.41  &1.41&  32\\
160&    -22.15  &2.89&  54\\
180&    -20.35  &1.30&  68\\
190&    -20.07  &1.26&  30\\
206&    -19.03  &1.90&  132\\
209&    -20.80  &1.83&  125\\
\hline                                          %inserts single line
\end{tabular}
\end{table}

The galaxies in the local Universe are observed to have gas-rich companions with H I content in 
the range $\sim$ 10$^6$ - 10$^8$ M$_{\odot}$ \citep{Sancisi2008}. We consider the case of gas-rich dwarf galaxy Leo P in the local Universe that have H I content $\sim$ 10$^6$ M$_{\odot}$ \citep{Giovanelli2013}. On interaction with a L$*$ ETG, this galaxy
 could supply fuel for less than million years with a modest star formation rate of 1 M$_{\odot}$/yr (assuming an efficient
 case of neutral-to-molecular hydrogen conversion). The fuel will exhaust more rapidly in the case of a star formation rate of the 18 M$_{\odot}$/yr observed in the extreme case of spiral galaxy tidal stripping event reported 
in this paper (Galaxy 180). This makes a cold gas supply, and the existence of blue ETGs becomes a transient phase in the evolution of a normal early-type galaxy and 
could explain the statistically low fraction [S09] of such galaxies in the local Universe. The extent of the interaction between the companion and the galaxy can be 
understood from deep optical or/and radio HI imaging and 
can be studied  for the stellar and gas exchange between the galaxies. The maintenance of the galaxy early-type morphology 
in a scenario of cold gas accretion from a companion followed by galaxy-scale star formation has to be addressed by detailed simulations. The fate of the interacting 
companion depends on the mass ratio between the galaxies and can either be a flyby or a major/minor merger. Recent literature points towards minor mergers 
as responsible for the increase in stellar mass in ETGs residing in low density environments \citep{Kaviraj_2014}. The galaxies without a detected companion in SDSS images 
could have gone through a minor merger phase and can explain the presence of jet features in the $g$ band images, blue colours, and enhanced star formation rates.\\

\section{Conclusions}
We presented here an analysis of  optical and near-UV imaging data of 55 star forming blue ETGs in order to 
understand the mechanism responsible for star formation.
We searched for signatures of recent interactions potentially triggering the star formation. Star forming blue ETGs show 
tidal trails and close proximity to neighbouring galaxies, which is evidence of
ongoing or recent interactions between galaxies. Out of 55 blue ETGs, 12 galaxies have very close blue 
companions of similar redshifts. Two of these galaxies are ongoing interacting systems with blue colours and high star formation rates. 
Two galaxies show a jet feature that could be due to tidal disruption of the companion galaxy. Blue ETGs appear to have high 
star formation rates (as high as 21 M$_{\odot}$/yr), which is only possible with an abundant cold gas supply from a reservoir. 
The recent or ongoing interaction with a 
gas-rich galaxy could be responsible for bringing molecular gas to an otherwise passively evolving red ETG. 
The gas supply could trigger 
the star formation, eventually creating a blue early-type galaxy. The galaxies with ongoing tidal interaction
 are blue and star forming, which implies that blue early-type galaxies can exist even when the companion is on flyby so does not need to end up in a merger.\\

\begin{acknowledgements}
%We thank Blesson Mathew for a careful reading of the manuscript and providing critical comments. 

KZ thanks the IIA for a summer intership programme when part of this work was carried out. KG is supported by an India-TMT postdoctoral fellowship. We thank H.C. Bhatt, Smitha Subramanian, and Blesson Mathew for carefully reading the manuscript and providing critical comments. The SDSS is managed by the Astrophysical Research Consortium for the Participating Institutions. The Participating Institutions are the American Museum 
of Natural History, Astrophysical Institute Potsdam, University of Basel, University of Cambridge, Case Western Reserve University, University of Chicago, Drexel University, Fermilab, the Institute for Advanced Study, the 
Japan Participation Group, the Johns Hopkins University, the Joint Institute for Nuclear Astrophysics, the Kavli Institute for Particle Astrophysics and Cosmology, the Korean Scientist Group, the Chinese Academy of 
Sciences (LAMOST), Los Alamos National Laboratory, the Max Planck Institute for Astronomy (MPIA), the Max Planck Institute for Astrophysics (MPA), New Mexico State University, Ohio State University, University of 
Pittsburgh, University of Portsmouth, Princeton University, the United States Naval Observatory, and the University of Washington. GALEX is operated for NASA by the California Institute of 
Technology under NASA contract NAS5-98034.  
\end{acknowledgements}

%-------------------------------------------------------------------


\begin{thebibliography}{99}
\bibitem[Abazajian et al.(2009)]{Abazajian_2009} Abazajian, K.~N., 
Adelman-McCarthy, J.~K., Ag{\"u}eros, M.~A., et al.\ 2009, \apjs, 182, 543 
\bibitem[{{Baldry} {et~al}\mbox{.}(2004){Baldry}, {Glazebrook}, {Brinkmann},
  {Ivezi{\'c}}, {Lupton}, {Nichol}, \& {Szalay}}]{Baldry_2004}
{Baldry} I.~K., {Glazebrook} K., {Brinkmann} J., {Ivezi{\'c}} {\v Z}., {Lupton}
  R.~H., {Nichol} R.~C., {Szalay} A.~S., 2004, \apj, 600, 681
\bibitem[Baldwin et al.(1981)]{Baldwin_1981} Baldwin, J.~A., 
Phillips, M.~M., \& Terlevich, R.\ 1981, \pasp, 93, 5 
\bibitem[Becker et al.(1995)]{Becker_95} Becker, R.~H., White, 
R.~L., \& Helfand, D.~J.\ 1995, \apj, 450, 559 
\bibitem[{{Bell} {et~al}\mbox{.}(2004){Bell}, {Wolf}, {Meisenheimer}, {Rix},
  {Borch}, {Dye}, {Kleinheinrich}, {Wisotzki}, \& {McIntosh}}]{Bell_2004}
{Bell} E.~F. {et~al.}, 2004, \apj, 608, 752
\bibitem[Brown et al.(2007)]{Brown_2007} Brown, M.~J.~I., Dey, A., 
Jannuzi, B.~T., et al.\ 2007, \apj, 654, 858 
\bibitem[Conroy et al.(2009)]{Conroy_2009} Conroy, C., Gunn, J.~E., 
\& White, M.\ 2009, \apj, 699, 486 
\bibitem[Dressler(1980)]{Dressler_1980} Dressler, A.\ 1980, \apj, 
236, 351 
\bibitem[{{Dressler} {et~al}\mbox{.}(1987){Dressler}, {Lynden-Bell},
  {Burstein}, {Davies}, {Faber}, {Terlevich}, \& {Wegner}}]{Dressler_1987}
{Dressler} A., {Lynden-Bell} D., {Burstein} D., {Davies} R.~L., {Faber} S.~M.,
  {Terlevich} R., {Wegner} G., 1987, \apj, 313, 42
\bibitem[{{Faber}(1973)}]{Faber_1973}
{Faber} S.~M., 1973, \apj, 179, 731
\bibitem[{{Faber} {et~al}\mbox{.}(2007){Faber}, {Willmer}, {Wolf}, {Koo},
  {Weiner}, {Newman}, {Im}, {Coil}, {Conroy}, {Cooper}, {Davis}, {Finkbeiner},
  {Gerke}, {Gebhardt}, {Groth}, {Guhathakurta}, {Harker}, {Kaiser}, {Kassin},
  {Kleinheinrich}, {Konidaris}, {Kron}, {Lin}, {Luppino}, {Madgwick},
  {Meisenheimer}, {Noeske}, {Phillips}, {Sarajedini}, {Schiavon}, {Simard},
  {Szalay}, {Vogt}, \& {Yan}}]{Faber_2007}
{Faber} S.~M. {et~al.}, 2007, \apj, 665, 265
\bibitem[Giovanelli et al.(2013)]{Giovanelli2013} Giovanelli, R., 
Haynes, M.~P., Adams, E.~A.~K., et al.\ 2013, \aj, 146, 15 
\bibitem[{{Kaviraj}(2010)}]{Kaviraj_2010}
{Kaviraj} S., 2010, \mnras, 408, 170
\bibitem[Kaviraj et al.(2011)]{Kaviraj_2011} Kaviraj, S., Tan, 
K.-M., Ellis, R.~S., \& Silk, J.\ 2011, \mnras, 411, 2148 
\bibitem[Kaviraj(2014)]{Kaviraj_2014} Kaviraj, S.\ 2014, \mnras, 
437, L41 
\bibitem[{{Komatsu} {et~al}\mbox{.}(2011){Komatsu}, {Smith}, {Dunkley},
  {Bennett}, {Gold}, {Hinshaw}, {Jarosik}, {Larson}, {Nolta}, {Page},
  {Spergel}, {Halpern}, {Hill}, {Kogut}, {Limon}, {Meyer}, {Odegard}, {Tucker},
  {Weiland}, {Wollack}, \& {Wright}}]{Komatsu_2011}
{Komatsu} E. {et~al.}, 2011, \apjs, 192, 18
\bibitem[Maraston et al.(2006)]{Maraston06} Maraston, C., Daddi, 
E., Renzini, A., et al.\ 2006, \apj, 652, 85 
\bibitem[Martin et al.(2005)]{Martin05} Martin, D.~C., Fanson, 
J., Schiminovich, D., et al.\ 2005, \apjl, 619, L1 
\bibitem[McIntosh et al.(2014)]{McIntosh14} McIntosh, D.~H., 
Wagner, C., Cooper, A., et al.\ 2014, \mnras, 442, 533 
\bibitem[{{Peng} {et~al}\mbox{.}(2002){Peng}, {Ho}, {Impey}, \&
  {Rix}}]{Peng_2002}
{Peng} C.~Y., {Ho} L.~C., {Impey} C.~D., {Rix} H.-W., 2002, \aj, 124, 266
\bibitem[{{Renzini}(2006)}]{Renzini06}
{Renzini} A., 2006, \araa, 44, 141
\bibitem[Sancisi et 
al.(2008)]{Sancisi2008} Sancisi, R., Fraternali, F., Oosterloo, T., \& van der Hulst, T.\ 2008, \aapr, 15, 189 
\bibitem[Schawinski et al.(2009)]{schawinski09} Schawinski, K., 
Lintott, C., Thomas, D., et al.\ 2009, \mnras, 396, 818 
\bibitem[Serra \& Trager(2007)]{Serra_2007} Serra, P., \& Trager, S.~C.\ 2007, \mnras, 374, 769 
\bibitem[Trager et al.(2000)]{Trager2000} Trager, S.~C., Faber, 
S.~M., Worthey, G., \& Gonz{\'a}lez, J.~J.\ 2000, \aj, 120, 165 
\bibitem[Visvanathan 
\& Sandage(1977)]{Visvanathan_77} Visvanathan, N., \& Sandage, A.\ 1977, \apj, 216, 214 
\bibitem[Young et al.(2011)]{Young_2011} Young, L.~M., Bureau, M., 
Davis, T.~A., et al.\ 2011, \mnras, 414, 940 
\bibitem[Young et al.(2014)]{Young_2014} Young, L.~M., Scott, N., 
Serra, P., et al.\ 2014, \mnras, 444, 3408 


\end{thebibliography}
\end{document}